\documentclass[aps,prl,reprint, groupedaddress]{revtex4-2}

\usepackage{siunitx}
\usepackage{graphicx}
\usepackage{amsmath,amssymb}
\usepackage{xcolor}

\newcommand{\be}{\begin{equation}}
\newcommand{\ee}{\end{equation}}
\newcommand{\bee}{\begin{eqnarray}}
\newcommand{\eee}{\end{eqnarray}}

\begin{document}


\title{Inertial Force Transmission in Dense Granular Flows}


\author{Matthew Macaulay}
\email[]{matthew.macaulay@sydney.edu.au}
\author{Pierre Rognon}
\email[]{pierre.rognon@sydney.edu.au}
\affiliation{School of Civil Engineering, The University of Sydney, Australia}


\date{\today}

\begin{abstract}
Dense granular flows are well described by several continuum models, however, their internal dynamics remain elusive. This study explores the contact force distributions in simulated steady and homogenous shear flows. Results demonstrate the existence of high magnitude contact forces in faster flows with stiffer grains. A proposed physical mechanism explains this rate-dependent force transmission. This analysis establishes a relation between contact forces and grain velocities, providing an entry point to unify a range of continuum models derived from either contact forces or grain velocity.
\end{abstract}


\maketitle

Granular flows transmit forces via a network of contacts between pairs of grains. Knowing the magnitude of these forces is key to predicting the dynamics of natural and industrial flows. First, contact forces are an elementary vector of momentum transport \cite{macaulay2020two}. As such, they control the effective viscosity of the material. Second, large contact forces can lead to grain comminution, which in turns affects the flow microstructure and its rheological behaviour \cite{marks2015mixture}. However, while several continuum models can describe dense granular flows \cite{midi2004dense,kamrin2012nonlocal}, the nature and the origin of their internal contact force distribution remain poorly understood. 

Pioneering works focusing on quasi-static packings revealed that (i) the mean contact force is controlled by the confining stress $P$ and the grain size $d$ as $Pd^2$ and (ii) some contacts are compressed to much higher levels \cite{radjai1998bimodal,majmudar2005contact}. A piecewise force distribution was manifest, including an exponentially decaying probability of finding contact forces with a magnitude above the mean. 

Contact forces in collisional flows are radically different as they arise from binary collisions between grains. In shear flows, Bagnold's scaling relates the typical impact velocity to the shear rate $\dot \gamma$ and the grain size $d$ as $\dot \gamma d$. Considering a simple elasto-inertial collision between grains of mass $m$ and contact stiffness $k$ provides an estimate for the maximum impact contact force of $\dot \gamma d \sqrt{km}$ \cite{macaulay2020viscosity}. This highlights a rate-dependence of the contact forces.

\begin{figure}
	\includegraphics[width=0.9\linewidth]{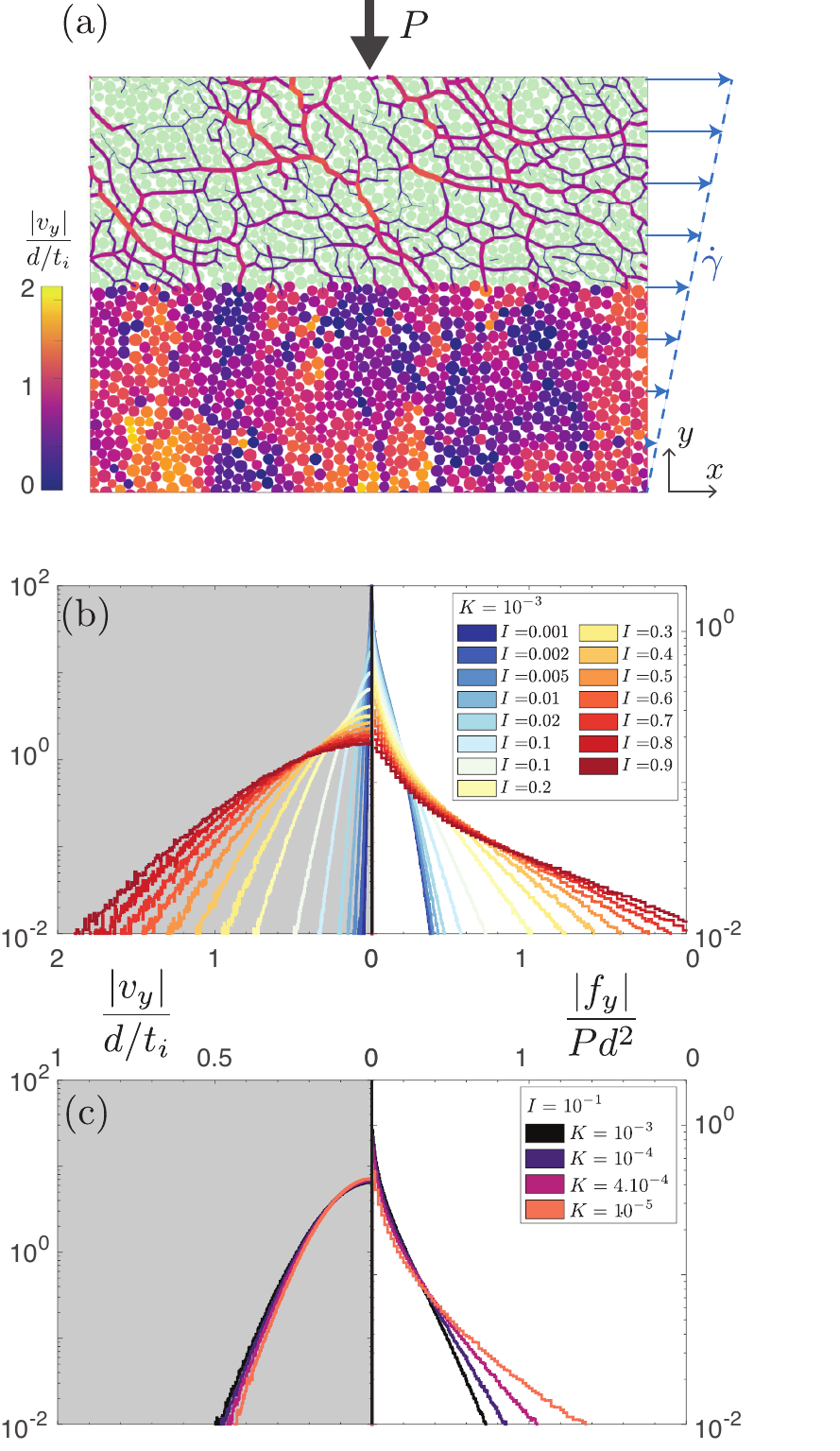}
	\caption{{\bf Force and velocity distributions.} (a) Plane shear under prescribed pressure $P$ and shear rate $\dot \gamma$, showing a snapshot of the contact forces (top half, represented by segments of width proportional to the force magnitude) and of the grain velocities (bottom half) in a flow with $I=0.01$ and $K=10^{-3}$. Movies available online show the evolution of contact forces and grain velocities within flows at different inertial number and softness. (b,c) PDF of grain velocity (left) and contact forces (right) for different inertial numbers (b) and different softness (c).
 \label{fig1}}
\end{figure}

\begin{figure}
	\includegraphics[width=0.9\linewidth]{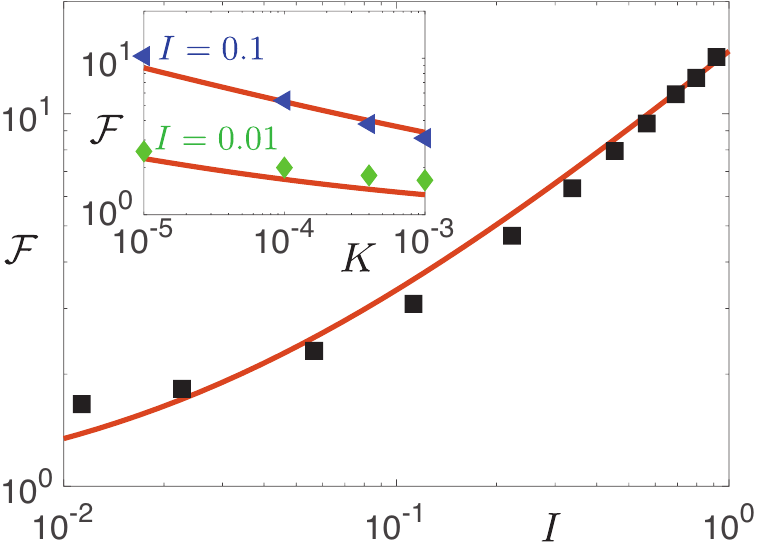}
	\caption{{\bf Contact force fluctuations $\mathcal{F}$} for different inertial numbers $I$ ($K=10^{-3}$, main) and different softness $K$ (inset). Symbols: numerical results; lines: proposed model (\ref{eq:pheno}). \label{fig2}} 
\end{figure}

In dense granular flows, binary collisions are seldom as grains typically experience multiple contacts. Their trajectories involve inertial reorganisation events whose dynamics is independent on, and much faster than the shear rate \cite{midi2004dense}. As a result, the grains' relative velocity deviates from Bagnold's scaling \cite{radjai2002turbulentlike,da2005rheophysics,degiuli2016phase}. The resulting distribution in grain velocity has been shown to control the process of shear-induced diffusion \cite{kharel2017vortices,macaulay2019shear}. However, how and whether it is related to contact forces is unknown.

In this Letter, we seek to establish a physically-based model that captures the force distribution in dense granular flows. In this aim, we have measured contact force distribution in simulated steady and homogenous shear flows and inferred a physical mechanism involving grain velocity fluctuations that is able to capture these observations.

\paragraph{Simulations.}

We used a discrete element method to simulate granular flows in a plane shear configuration (figure \ref{fig1}a). The shear cell is bi-dimensional. Both the normal stress $P$ and the shear rate $\dot \gamma$ are prescribed using periodic boundary conditions to avoid the flow heterogeneities that solid walls would induce \cite{rognon2015long}. The granular material is comprised of $10^4$ grains of mean diameter $d$, mass $m$ and density $\rho$. A uniform polydispersity of $d\pm20\%$ is introduced on grain's diameter to prevent shear-induced crystallisation. Grains interact via pairwise contacts including a normal elasto-dissipative force and a tangential friction. The supplemental material further details these granular interactions and the simulation procedure. 
Contact parameters include a coefficient of restitution and a coefficient of friction, which are both set to $0.5$. In contrast, different values of the contact normal stiffness $k$ will be considered. The effect of these contact parameters on the granular rheology is discussed in \cite{da2005rheophysics}. 

The following analysis focuses on steady and homogenous flows obtained with this configuration while prescribing two dimensionless numbers $I$ and $K$:

\be
I = \dot \gamma t_i; \quad t_i = d\sqrt{\frac{\rho}{P}}; \quad K = \frac{t^2_c}{t^2_i} = \frac{Pd}{k}.
\ee

\noindent The \textit{inertial} number $I$ compares the shear time $\dot \gamma^{-1}$ to the inertial time $t_i$. $t_i$ measures a characteristic time for a grain of mass $m$ initially at rest to move over a distance $d$ under the action of a force $Pd^2$ \cite{midi2004dense,da2005rheophysics}. The \textit{softness} number $K$ measures the elastic deformation of a grain experiencing a compressive contact force of magnitude $Pd^2$; equivalently, it compares the inertial time $t_i$ to an elasto-inertial collision time $t_c = d \sqrt{\frac{\rho d}{k}}$. In the following, results cover the ranges $10^{-3} \leqslant I \leqslant .9$ and $10^{-5} \leqslant K \leqslant10^{-3}$, which correspond to the dense flow regime and the rigid limit, characterised by $t_c<t_i<\dot \gamma^{-1}$ \cite{midi2004dense,da2005rheophysics}.

\begin{figure}
	\includegraphics[width=\linewidth]{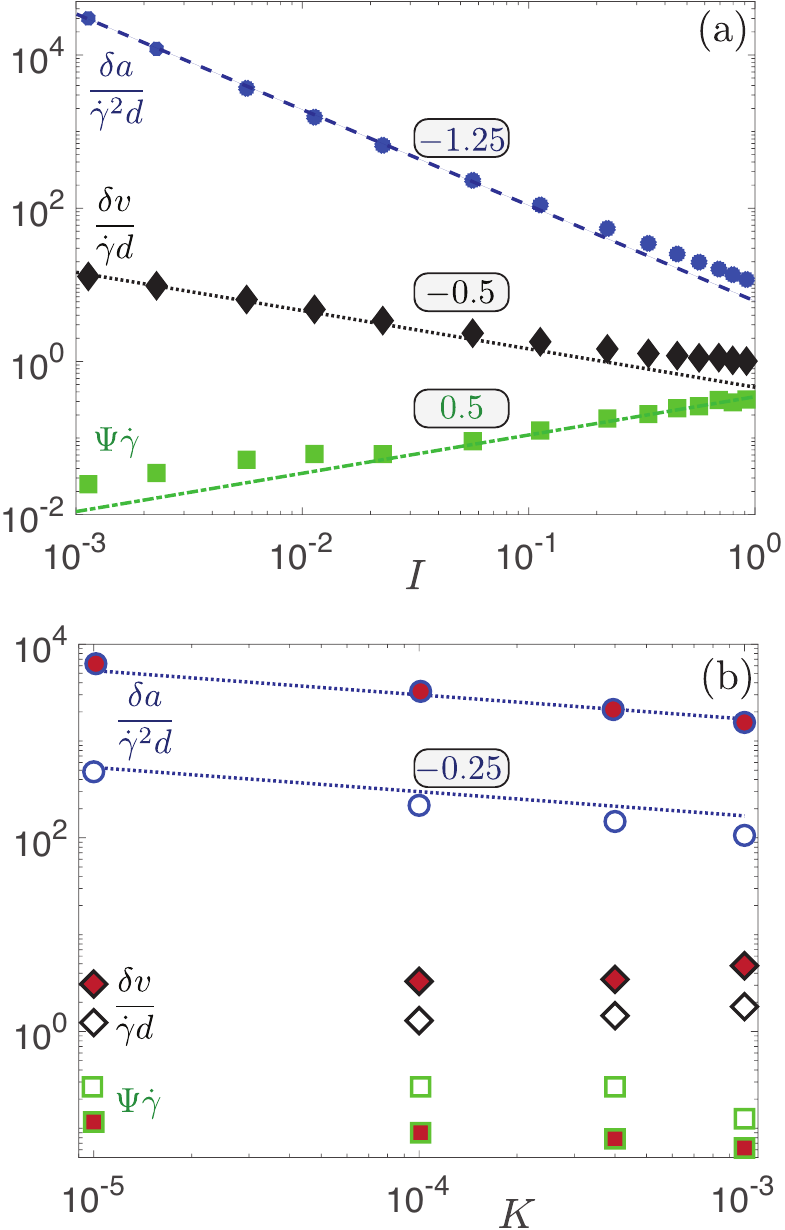}
	\caption{{\bf Internal kinematic variables: } Velocity fluctuation $\delta v$, acceleration fluctuation $\delta a$ and velocity memory time $\Psi$ measured in flows with: (a) different inertial numbers ($K=10^{-3}$); and (b) different softness $K$ for $I = 10^{-2}$ (filled symbols) and $I = 10^{-1}$ (open symbols). Symbols: numerical results; lines: proposed models in (\ref{eq:AKI}), (\ref{eq:vI}) and (\ref{eq:psiI}). Numbers summarise the power-law exponents. \label{fig3}}
\end{figure}

\paragraph{Phenomenological scaling for force fluctuation.}
First, we measured the component of the contact forces in the direction transverse to the shear, $|f_y|$, to determine their probability density function (PDF). Taking advantage of the homogeneity and steadiness of the flows, these were measured considering all grains in the cell and $300$ sampled times spread evenly over $15$ shear deformations. Figures~\ref{fig1}b,c show that the force distributions are affected by both the inertial number $I$ and the softness $K$. To characterise the width of these distributions, we measured their standard deviation $\delta f$ which we refer to as force fluctuation. 
Figure \ref{fig2} shows the combined influence of $I$ and $K$ on the normalised force fluctuation $\mathcal{F} =\frac{\delta f}{Pd^2}$. $\delta f$ converges to $Pd^2$ in the quasi-static limit $I\to0$, which is consistent with the findings in \cite{radjai1998bimodal,majmudar2005contact}. In contrast, it becomes significantly larger at high inertial numbers and for stiffer grains: $\delta f$ reaches values up to one order of magnitude larger than $Pd^2$ in the range of parameters that have been explored. However, these results do not immediately provide evidence for a relation between $\delta f$ and the dimensionless numbers $I$ and $K$.

To identify this relation, we use the micro-inertia theory introduced in \cite{macaulay2020two} that relates contact force fluctuation to grain acceleration. The rationale for this theory is that multiple contact forces acting on a grain may be partially balanced. Their balanced part does not induce any acceleration and scales with $Pd^2$. Their unbalanced part drives some grain acceleration according to Newton's law of motion. This ultimately yields a linear relation between force and acceleration fluctuations $\delta f \approx Pd^2 +m\delta a$. Here, the acceleration fluctuation is defined as the standard deviation of the distribution in grain acceleration in the $y$-direction. A dimensionless formulation of this relation is:

\be \label{eq:FA}
\mathcal{F} \approx 1 + \mathcal {A},
\ee

\noindent where $\mathcal {A} = \frac{\delta a}{d/t_i^2}$ is the normalised acceleration fluctuation. The supplemental section shows that this relation is valid for flows performed at different inertial numbers $I$ and with different softness $K$. Thanks to this relation, establishing how the force fluctuation varies with $I$ and $K$ may be achieved by finding how the acceleration fluctuation varies with $I$ and $K$. 

Figure \ref{fig3} shows that the acceleration fluctuation increases with higher inertial number and stiffer grains, seemingly with a power law: 

\be \label{eq:AKI}
\frac{\delta a}{d\dot \gamma^{2}} \propto K^{n_{aK}} I^{n_{aI}}.
\ee 

\noindent Results are consistent with exponents $n_{aK} = -0.25$ and $n_{aI} = -1.25$. Introducing this scaling into the force-acceleration relation (\ref{eq:FA}) leads to the following phenomenological scaling law for the force fluctuations: 

\be \label{eq:pheno}
\mathcal{F} \approx 1 + \alpha K^{-0.25} I^{0.75}.
\ee

\noindent Figure \ref{fig2} shows that this relation reasonably captures the force fluctuation in the explored range of $I$ and $K$ with $\alpha =2.5$ as a fitting parameter.

\begin{figure}
	\includegraphics[width=\linewidth]{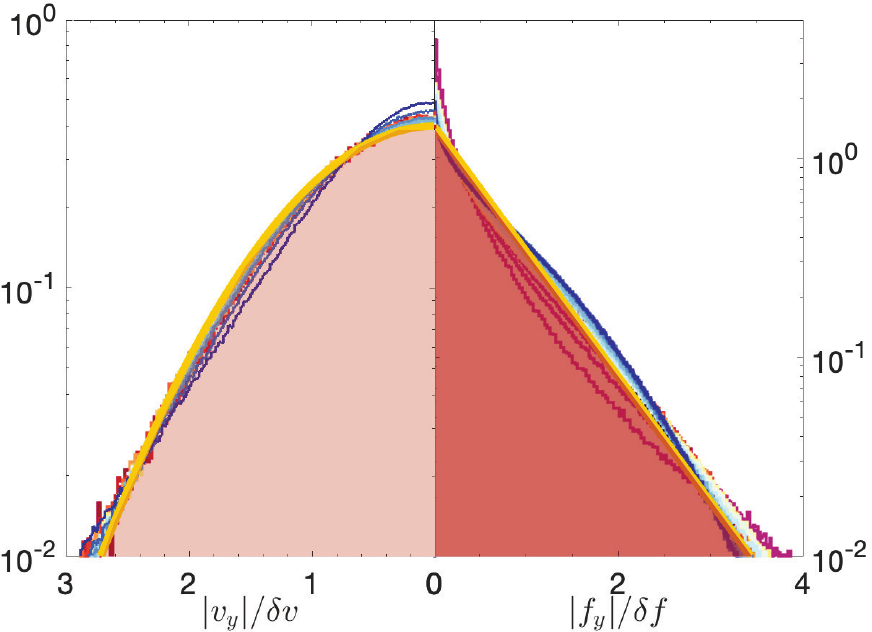}
	\caption{{\bf Rescaling of force and velocity distribution}. Lines are the PDF presented in figure \ref{fig1}b,c, rescaling grain velocity and contact forces by $\delta v$ and $\delta f$ defined in the models (\ref{eq:vI}) and (\ref{eq:pheno}), respectively. Filled areas show a normal distribution $ (2\pi)^{-\frac{1}{2}} e^{-\frac{1}{2}\left(|v_y|/\delta v \right)^2}$ (left) and an exponential distribution $e^{-\beta \frac{|f_y|}{ \delta f} } $ (right) with $\beta=1.4$ found as best fitting the data. 
	\label{fig4}}
\end{figure}

\paragraph{Physical process relating velocity and force fluctuations.}

To identify a physical process at the origin of this phenomenological law, we now seek to relate the acceleration fluctuation to other kinematic quantities including grain velocities and their time persistence. 
 
Figures \ref{fig1}b,c show that the grain vertical velocity distribution is affected by the inertial number. However, unlike contact forces, the velocity distribution is not dependent on the softness. Figure \ref{fig3} reports the standard deviation of these distributions, which we refer to as velocity fluctuations $\delta v$. Results suggest the following power law:

 \be \label{eq:vI}
\frac{\delta v}{\dot \gamma d } \approx 0.5 I^{n_{v}}.
\ee

\noindent As previously proposed \cite{da2005rheophysics,kharel2017vortices,macaulay2019shear}, they are consistent with a power exponent $n_{v}=-0.5$. They further point out the negligible influence of the softness $K$. 
 
Figure \ref{fig3} also reports the velocity auto-correlation time $\Psi$, defined as $\Psi = \int_{t=0}^{\infty} \psi(t)dt$, where $\psi(t) = \langle v_y(t_0) v_y(t_0+t) \rangle / \langle v_y(t_0)^2 \rangle$ is the velocity auto-correlation function and the angle brackets denote the average operator on all grains and all reference times $t_0$. We refer to $\Psi$ as velocity persistence time, as it measures a characteristic period of time during which a grain velocity $v_y$ is sustained. Results suggest the following power law:

 \be \label{eq:psiI}
\Psi \dot \gamma \approx 0.35 I^{n_{\Psi }}.
\ee

\noindent As previously reported \cite{degiuli2016phase,kharel2017vortices,macaulay2019shear}, they are consistent with a power exponent $n_{\Psi} = 0.5$. They further highlight the near-independence of the velocity memory time on the softness $K$. 

From these two observations on the velocity fluctuation and their persistence, we infer a simplified scenario for a typical grain motion in dense granular flows. Like a random walk, this scenario involves grains moving over steps of length $\delta v \Psi$ at a constant velocity $\delta v$ before changing direction. Accordingly, the acceleration during the step is null. As for collisional flows, we postulate that the acceleration during the change of direction is of the order of $\delta v/t_c$: this reflects a change in grain velocity of $\pm \delta v$ in a period of time driven by the elasto-inertial collision time $t_c$. 

The characteristic mean square acceleration during such a cycle is (considering than $t_c\ll \Psi$): $\langle a^2 \rangle = \frac{\big(\frac{\delta v}{t_c}\big)^2 t_c+0}{\Psi} = \frac{\delta v^2}{t_c \Psi}$. This predicts the following relation between velocity fluctuation and acceleration fluctuation $\delta a \equiv \langle a^2 \rangle^{0.5}$: 

\be \label{eq:av}
\delta a \propto \frac{\delta v}{\sqrt{t_c \Psi}}.
\ee

\noindent Introducing this relation into the force-acceleration relation in (\ref{eq:FA}) leads to the following relation between the force and velocity fluctuations:

\be \label{eq:fv}
\delta f - Pd^2 \propto m\frac{\delta v}{\sqrt{t_c \Psi}}.
\ee

\noindent Introducing in (\ref{eq:fv}) the scalings for the velocity fluctuation and their lifetime in (\ref{eq:vI}) and (\ref{eq:psiI}) leads to the following scaling law for the force fluctuation expressed in terms of the inertial number and the softness $K$: $\mathcal{F} - 1 \propto K^{-\frac{1}{4}} I^{\frac{5}{4}}$, which is consistent with the phenomenological model (\ref{eq:pheno}). The supplemental section further assesses the validity of this scenario by comparing the measured $\delta a$ and $\delta f$ to the predictions of Eqs (\ref{eq:av}) and (\ref{eq:fv}), respectively. 

Finally, figure \ref{fig4} shows that the distribution of contact forces collapse onto a single exponential distribution when rescaled by the force fluctuation defined by (\ref{eq:pheno}). This distribution is similar to that found in quasi-static packings \cite{radjai1998bimodal,majmudar2005contact}. Here, we find how its width depends on the inertial number and grain softness in dense flows. Similarly, figure \ref{fig4} shows that grain velocity distributions all collapse onto a single normal distribution when rescaled by the velocity fluctuation modelled by (\ref{eq:vI}). This further confirms that, unlike contact forces, grain kinematic is independent on their softness. 

 \paragraph{Conclusions.} This study highlights a significant rate-dependence of contact forces in dense granular flows. It reveals the existence of forces much larger than the characteristic scale $Pd^2$ with an exponential distribution, and introduces a model (\ref{eq:pheno}) to capture the width of this distribution as a function of the inertial number and grain softness. These results can readily be used to foresee flow conditions likely to lead to grain breakage or abrasive wear induced by extreme contact forces. 
 
The random-walk mechanism introduced to explain this rate-dependence of the force distribution highlights a general relationship between contact forces and grain velocity fluctuation (\ref{eq:fv}). This relationship is an entry point to help reconcile two distinct existing approaches to understand the constitutive behaviour of dense granular flows that are based on either contact forces \cite{pouliquen2009non,azema2014internal,macaulay2020two} or grain velocity fluctuations \cite{zhang2017microscopic,kim2020,gaume2020}. Lastly, we anticipate that analogous physical processes could govern the force distribution in other soft materials such as suspensions, foams and emulsions, which would involve each material's specific mode of interaction working via an interstitial liquid.

%

\end{document}